\documentclass[aps,pra,showpacs,twocolumn,superscriptaddress,floatfix]{revtex4-2}
\usepackage{graphicx}
\usepackage{dcolumn}
\usepackage{bm}
\usepackage{hyperref}
\usepackage{comment}

\begin{document}

\preprint{APS/123-QED}

\title{Measuring dark state number in the Tavis-Cummings model}

\author{L. Theerthagiri}
 
\author{Rajesh Narayanan}
\affiliation{
 Department of Physics, Indian Institute of Technology Madras, Chennai 600036, India}

\author{R. Ganesh}
\email{r.ganesh@brocku.ca}
\affiliation{
Department of Physics, Brock University, St. Catharines, Ontario L2S 3A1, Canada
}

\date{\today}

\begin{abstract}
Quantum mechanics allows for light-matter setups that hold excitations without releasing them as light. Arising from destructive interference processes, they are best seen in a Tavis-Cummings-like setup where two-level atoms (or qubits) are placed within a lossy cavity. If the system is initialized with some qubits excited and some in the ground state, there is a non-zero probability that no photons will be emitted. This can be framed as a Stern-Gerlach measurement, with a detector to measure if one or more photons leave the cavity. If no photons are detected, the qubits collapse onto a dark state. This can be viewed as heralding of a dark state based on zero photon detection. 
Building upon this idea, we propose a protocol to measure the number of independent dark states. Moreover, we show that this quantity is robust to arbitrary levels of disorder in the qubit-photon coupling constants. We then discuss a phase transition where the number of dark states plays the role of an order parameter. This provides an exciting example of a phase transition that is completely insensitive to disorder. 
\end{abstract}

%\keywords{Suggested keywords}

\maketitle

\section{Introduction}
Dark states are long-lived quantum states that hold excitations without releasing them as light\cite{holzinger,Wang2020,VillasBoas2025}. They are closely related to the phenomenon of subradiance. They have the potential to serve as quantum memories that can store information. Motivated by this technological promise, they have been explored in several theoretical and experimental studies. The underlying notions may be traced back to a seminal thought experiment of Dicke from 1954 with two emitters\cite{Dicke1954}. This was experimentally realized in 2014 using superconducting qubits in a microwave cavity\cite{Mlynek2014}. A key feature of this experiment was the lossy-cavity limit, where the time scale for a qubit-photon coupling is much longer than that for a photon to leave the cavity. This allows an emitted photon to `immediately' leave the cavity be detected outside.

Inspired by the lossy-cavity setup, two of the current authors have proposed Stern-Gerlach-like protocols to synthesize dark states\cite{Giri2017}. A measurement is performed by a photon detector placed outside the lossy mirror. A null result (zero photons detected) is a possible outcome. As with any quantum measurement, the null result will collapse the qubits within onto a dark state. The resulting state is entangled, expressible as a resonating valence bond wavefunction. In this article, we build on these ideas to (i) develop a protocol to measure $N_{dark}$, the number of dark states, (ii) establish that $N_{dark}$ is robust to arbitrary levels of disorder, and  (iii) describe a phase transition where $N_{dark}$ serves as the order parameter.

A crucial ingredient in our arguments is the role of a zero-outcome in photon measurement\cite{Nunn2021,Cryer2025,Clarke2025}. It is well known that detection of a single photon can herald a specific quantum state in a second channel\cite{Migdall2002,Pittman2002}. Studies have explored the possibility of a zero-measurement doing the same\cite{Giri2017,Cryer2025,Clarke2025}.
Here, we view a cavity-QED setup as two channels -- one of qubits and one of photons. Non-detection in the photon channel (a zero-measurement) can place the qubits in a specific state. We exploit this notion to generate and even count dark states.

The Tavis-Cummings model\cite{TC1968,wierzchucka} has been realized in ultracold atomic gases, circuit-QED setups\cite{Fink2009,Feng2015,Wang2020,filipp}, cavity-QED\cite{Kim,Tolazzi,sundar}, nitrogen-vacancy centres\cite{Kubo2010,Roopini2020}, cold atom gases\cite{Johnson2019,roof,ivor,guerin} etc. Any experimental setup will involve some level of disorder. In the context of the Tavis-Cummings model, disorder can enter in diagonal or off-diagonal terms. The former corresponds to randomness in qubit energies (frequencies) while the latter pertains to the coupling of each qubit to the photon mode. We restrict our attention to off-diagonal disorder below.

The impact of disorder on phases and phase transitions of quantum systems has been extensively studied \cite{vojta2013phases,Vojta19}.
Quenched or ``frozen-in'' disorder can have drastic effects. In low dimensions ($d\le 2$) \cite{aizenman_wehr,hui}, quenched disorder rounds first order transitions. It may destroy a putative critical point via smearing \cite{jose_smeared_prb,jose_smeared_prl,Kaur_2024}. It may produce entirely new critical behaviour, e.g., an 
Infinite Randomness Fixed Point (IRFP) with attendant Griffiths phases\cite{Fisher92,Jose_2007,VojtaKotabageHoyos09}. 
These ideas have been experimentally explored in superconductors\cite{xing2015quantum,shen2016,lewellyn2019infinite,zhang2019quantum,Kaur_2024} and magnets \cite{almut1, almut2}. In the context of light-matter interactions, there have been relatively few studies on the effects of disorder on phase transitions\cite{Goto2008,Kirton2017,Das2024}. 
Here, we use the disorder-insensitive character of $N_{dark}$ to devise an intriguing phase transition.  Qualitatively, the transition occurs between a bright state (where $N_{dark}$ vanishes) and a dark state. Remarkably, the phase transition here is unchanged with any amount of disorder. This gives an interesting example of a phase transition that is completely insensitive to disorder.

\section{Dark states in the uniform limit}
\label{sec.clean}
In the interest of completeness, we first briefly summarize the problem in the absence of disorder. The results here have been demonstrated in Refs.~\cite{Giri2017} and \cite{Giri_Scipost}. We consider the Tavis-Cummings model with $N$ qubits coupled to a single photon mode. Within the rotating wave approximation, the Hamiltonian is given by
\begin{eqnarray}
    H_{TC} = \omega (a^\dagger a +  S_{(tot.)}^z) + g \left\{ S_{(tot.)}^+a + S_{(tot.)}^-a^\dagger \right\}, 
    \label{eq.Huniform}
\end{eqnarray}
where $S_{(tot.)}^{\alpha} = \sum_{j=1}^N S_{j}^\alpha$ represents a collective spin operator ($\alpha = z, +,-$). The qubits are assumed to be resonant with the photon mode. The state of the qubits can be expressed using total angular momentum variables. For example, a state where all qubits are excited can be characterized as $(S_{(tot.)}=N/2;~~m_{(tot.)}=N/2)$. In this language, the photon-emission process encoded by the term $\{{S}_{(tot.)}^{-} a^\dagger\}$ preserves the quantum number $S_{(tot.)}$, but lowers $m_{(tot.)}$ by unity. We define a dark state as one that cannot emit a photon. From the allowed values of angular momentum, any stationary state with $(S_{(tot.)} =\lambda;~m_{(tot.)}=-\lambda)$ can be labelled as dark. Such a state cannot emit a photon as $m_{(tot.)}$ is at the minimum possible value and cannot be lowered further.

A crucial result emerges from the quantum addition-of-angular-momenta rules. The number of dark states with $s$ qubits in the excited state and $(N-s)$ qubits in the ground state is given by a simple formula. These states are characterized $m_{(tot.)}=-S_{(tot.)}=s-N/2$. The number of such states is 
\begin{eqnarray}
N_{dark,N,s} = \left\{ \begin{array}{cc}
0, & \mathrm{if}~s>N/2,\\
\left( \begin{array}{c} N \\ s \end{array}\right) - \left( \begin{array}{c} N \\ s-1 \end{array}\right), & \mathrm{if}~s\leq N/2.\\
\end{array}
\right.~~~~~
\label{eq.Ndark_clean}
\end{eqnarray}
A derivation of this expression is given in Appendix.~\ref{app.Ndark_clean}.

To realize such states in experiments, a Stern-Gerlach-like protocol can be employed\cite{Giri2017}. A direct-product initial state is placed within a lossy cavity. The state is prepared with $s$ qubits excited and $(N-s)$ qubits in the ground state, corresponding to $m_{(tot.)}=s-N/2$. A photon detector is placed outside at the lossy mirror. If one or more photons is detected, the run is discarded and the protocol repeated. Non-detection of photons (over a time scale set by the qubit-photon coupling) collapses the qubits onto a dark state. The precise wavefunction is obtained by projecting the initial state onto the sector with $S_{(tot.)}=(-)(s- N/2)$.

Ref.~\onlinecite{Giri_Scipost} describes a bright-to-dark phase transition that emerges upon taking the total number of qubits $N$ to infinity. The tuning parameter is $s$, equivalent to the polarization in the initial state. The order parameter is the number of photons emitted --- see Fig.~7 in Ref.~\onlinecite{Giri_Scipost}. Below, in Sec.~\ref{sec.transition}, we describe a modified version of this phase transition that remains robust to arbitrary levels of disorder.

\section{Disorder in light-matter coupling}
We now introduce disorder in the Tavis-Cummings model, with the Hamiltonian,
\begin{eqnarray}
    H_{dis.TC} = \omega(a^\dagger a + S_{(tot.)}^{z})+\sum_{j=1}^N  \left\{ g_j^* S_{j}^{+}a + g_j S_{j}^{-}a^\dagger \right\}\!,~~~ 
    \label{eq.disorderH}
\end{eqnarray}
where $g_j$'s are independent, possibly non-uniform, coupling constants at each qubit. 
Note that we allow for disorder in the qubit-photon coupling constants, but not in the qubit frequencies.  
Experimental realizations of the Tavis-Cummings model have been achieved using superconducting qubits, gases of sodium atoms, etc. Disorder of the type encoded in Eq.~\ref{eq.disorderH} above is inevitable in such realizations. The position of each qubit within the cavity will affect the strength of the electric field as well as the local phase of the photon mode. In turn, these will affect the magnitude and phase of the coupling $g_j$. 

\section{Robustness of dark states with disorder}
\label{sec.robust}

We define dark states as long-lived states that carry excitations but do not emit photons. The requirements for such a state are (i) it must be stationary, i.e., it must be an eigenstate of the Hamiltonian, (ii) it must not contain photons as any photons will eventually be emitted, and (iii) it must be annihilated by 
\begin{eqnarray}
    \hat{O}_{dark} = \sum_j g_j S_{j,-},
    \label{eq.Odarkdef}
\end{eqnarray}
the operator that appears alongside $a^\dagger$ in the Hamiltonian of Eq.~\ref{eq.disorderH}. A state that satisfies these requirements may possibly absorb a photon if one is available from the environment, e.g., see discussion on subradiant states in Ref.~\onlinecite{Wang2020}. We persist with our definition of a dark state as one that does emit, as photon absorption is highly unlikely without external pumping. 

The action of $\hat{O}_{dark}$ can be separated into various sectors,
\begin{eqnarray}
 \hat{O}_{dark} = \sum_{s=0}^{N} \hat{O}_s,  
\end{eqnarray}
where $s$ represents the number of qubits that are initially excited. As $\hat{O}_{dark}$ is a linear combination of spin-lowering operators, the final state after the action of $\hat{O}_{dark}$ has $s-1$ qubits excited. We note that each $s$ corresponds to a different sector of the Hilbert space. 

We may write $\hat{O}_s$ as a matrix of dimensions $\left(\begin{array}{c} N\\ s -1 \end{array}\right)\times \left(\begin{array}{c} N\\ s\end{array}\right)$. This can be viewed as a linear transformation from the $s$-sector to the $(s-1)$-sector. Based on arguments presented in Appendix.~\ref{app.rank}, we find the rank of $\hat{O}_s $ to be
\begin{eqnarray}
    \mathrm{Rank}\{ \hat{O}_s \} = 
    \left\{ \begin{array}{cc}
\left( \begin{array}{c} N \\ s \end{array}\right), & \mathrm{if}~s>N/2,\\
 \left( \begin{array}{c} N \\ s-1 \end{array}\right), & \mathrm{if}~s\leq N/2.\\
\end{array}
\right.~~~~~
\label{eq.rank}
\end{eqnarray}
Using the rank-nullity theorem\cite{friedberg2014linear}, we have
\begin{eqnarray}
  \mathrm{Nullity}\{ \hat{O}_s \} =  
    \left\{ \begin{array}{cc}
0, & \mathrm{if}~s>N/2,\\
 \left( \begin{array}{c} N \\ s \end{array}\right)-\left( \begin{array}{c} N \\ s-1 \end{array}\right), & \mathrm{if}~s\leq N/2.\\
\end{array}
\right.~~~~~
\label{eq.nullity}
\end{eqnarray}
Here, nullity represents the number of eigenstates of $\hat{O}_s$ with eigenvalue zero. 
Such states (with no photons in the cavity mode) satisfy all requirements for a dark state: 
(i) they are eigenstates of the Hamiltonian in Eq.~\ref{eq.disorderH}, as they have a well-defined $S_{(tot.)}^z$ quantum number, (ii) they are annihilated by the $\hat{O}_s$ operator which accompanies $a^\dagger$, and (iii) they are annihilated by the operator $a$ as there are no photons. This guarantees a long-lived state that does not emit photons.

The expression in Eq.~\ref{eq.nullity} above thus yields the number of independent dark states. Note that this result is precisely the same as Eq.~\ref{eq.Ndark_clean}, which describes the disorder-free uniform limit.

Remarkably, this result holds for arbitrary values of $g_j$'s, as long as all $g_j$'s are non-zero. In the section below, we demonstrate the validity of this assertion by explicitly constructing dark states for a certain sector.

\section{Dark states with a single excitation}

We now present analytic forms for dark states in the single-excitation sector, i.e., with $s=1$. In this sector, $\hat{O}_{dark}$ of Eq.~\ref{eq.Odarkdef} takes the form of a $1\times N$ matrix,
\begin{eqnarray}
\hat{O}_{dark}^{(s=1)} = \left(
\begin{array}{ccccc} 
g_1 & g_2 & \ldots & g_{N-1} & g_N
\end{array}\right).
\label{eq.OdarkL1}
\end{eqnarray}
Each entry is obtained as $ \langle \downarrow \ldots \downarrow \vert  \sum_j g_j S_{j,-} \vert e_j\rangle$, where $\vert e_j \rangle = \vert \downarrow_1 \ldots \uparrow_j \ldots \downarrow_N\rangle$ represents a single-excitation-state where the $j^{\mathrm{th}}$ qubit alone is excited. 

From Eq.~\ref{eq.nullity}, we have $N-1$ dark states in the $(s=1)$ sector. The simple structure of $\hat{O}_{dark}$ allows us to write analytic expressions for the wavefunctions of these dark-states. We write
\begin{equation}
    \vert d_j\rangle=
    \frac{\vert g_j \vert}{\sqrt{\vert g_j\vert^2+\vert g_N\vert^2}}
    \biggl\{ -\left(\frac{g_N}{g_{j}}\right)
\vert e_j \rangle
+ \vert e_N \rangle
    \biggr\}\otimes \vert n_{ph}=0\rangle,\quad 
\end{equation}
where $j=1,2,\dots N-1$. Here, $\vert n_{ph}=0\rangle$ denotes that the photon sector is empty, with no photons in the cavity mode. We readily see that each $\vert d_j \rangle$ is annihilated by $\hat{O}_{dark}^{(s=1)}$.  Moreover, it is an eigenstate of the ${S}_{(tot.)}^z$ operator. As a result, each $\vert d_j \rangle$ is an eigenstate of the Hamiltonian in Eq.~\ref{eq.disorderH}. It follows that $\vert d_j\rangle$'s represent long-lived states that contain a single excitation, but do not release it as light. While $\vert d_j\rangle$ states are not orthogonal to one another, they can be easily seen to be linearly independent. We thus have an $(N-1)$-dimensional dark subspace. Expressions for dark states in the single-excitation sector can also be found in Ref.~\onlinecite{Zeb2022}.

\section{Nullity as an observable}
\label{sec.nullity}

We now describe a protocol to experimentally measure the number of dark states for a given $s$. We assume a light-matter setup that realizes the Tavis-Cummings model with or without disorder. Qubits are placed within a cavity with a lossy mirror. In the lossy-cavity limit\cite{Mlynek2014}, the time-scale for a photon to leave the cavity is much shorter than that associated with qubit-photon coupling. If a photon is emitted by the qubits, it `immediately' leaves the cavity via the lossy mirror. We propose the following protocol:

\begin{figure}
\includegraphics[width=3.4in]{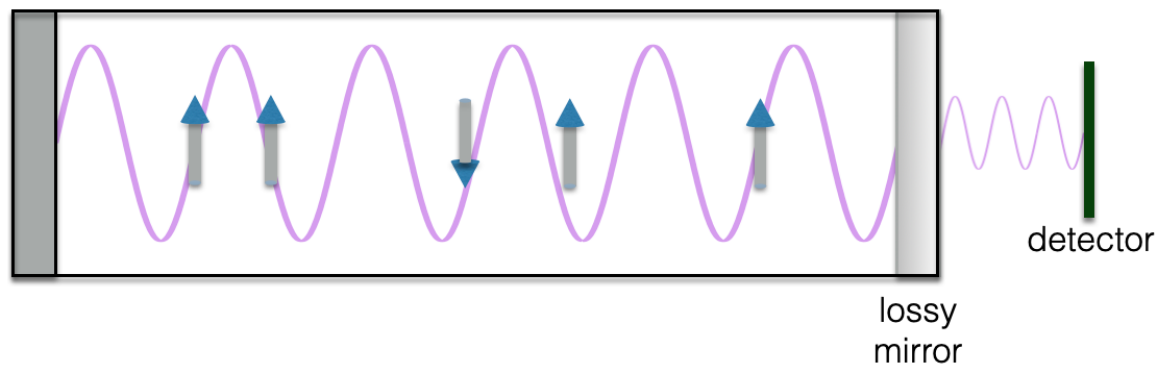}
\caption{Schematic setup required for protocol to measure number of dark states. Qubits are placed within a lossy cavity, with a photon detector outside. We may have variations in the amplitude of the photon mode from one qubit position to the next. This can produce randomness in the qubit-photon coupling constants, which does not affect the outcome.}
\label{fig.setup}
\end{figure}

\begin{itemize}
\item An initial direct product state is synthesized where $s$ qubits are excited and $N-s$ are in the ground state. This initial state is placed within the cavity.

\item A photon detector (assumed to be 100\% efficient), placed outside the lossy mirror, is monitored. If the detector clicks, it signals the emission of one or more photons. This is recorded as an emission event. 

\item If the detector does not click over a sufficiently long time (set by the weakest qubit-photon coupling), we conclude that the initial state has collapsed onto a dark state. This is recorded as a null-emission event. 

\item The above steps are repeated several times to obtain a reliable estimate of the probability of null emission. 

\item The entire process is repeated over all rearrangements of the initial state, i.e., over all possible arrangements of $N$ qubits where $s$ are excited. Upon \textit{adding} the probability of null emission from each rearrangement of $N$ qubits, we obtain the number of dark states. 
\end{itemize}

In mathematical terms, the outcome of this protocol can be described as follows. We first define $\hat{\mathcal{P}}_{dark,N,s}$, an operator that projects onto the dark subspace --- the space spanned by the null eigenstates of $\hat{O}_s$. We have 
\begin{eqnarray}
 \hat{\mathcal{P}}_{dark,N,s} = \sum_{j=1}^{N_{dark,N,s}} 
 \vert d_j \rangle \langle d_j \vert,
\end{eqnarray}
where $\vert d_j\rangle$ runs over all dark states in the $s$-sector. For a given initial state $\vert \psi_{init.}
\rangle$, the probability of null emission is given by $
\langle \psi_{init.} \vert      \hat{\mathcal{P}}_{dark,N,s}
\vert \psi_{init.} \rangle$. 

We next define a permutation operator, $\hat{\pi}_{k,N,s}$. Here, $k$ runs over the $\left(
\begin{array}{c}
N\\
s
\end{array}
\right)$ ways to select $s$ qubits out of $N$. The $s$-sector is spanned by $\left(
\begin{array}{c}
N\\
s
\end{array}
\right)$ basis states that can be expressed as
\begin{eqnarray}
 \vert \psi_k \rangle  =   \hat{\pi}_{k,N,s} \vert \psi_{ref.} \rangle,
\end{eqnarray}
where $\vert \psi_{ref.} \rangle$ is a reference state, say with the first $s$ qubits excited while the rest are in the ground state. The $\vert \psi_k\rangle$'s represent all rearrangements of the excited qubits.

The final outcome of the protocol can be expressed as
\begin{eqnarray}
\nonumber D(s) &=& \sum_{k} \Big\{\langle 
\psi_{ref.} \vert 
\hat{\pi}_{k,N,s}\Big\}
\hat{\mathcal{P}}_{dark,N,s}
\Big\{
\hat{\pi}_{k,N,s}
\vert \psi_{ref.} \rangle\Big\}~~~~\\
\nonumber &=& \sum_{k} \langle 
\psi_{k} \vert 
\hat{\mathcal{P}}_{dark,N,s}
\vert \psi_{k} \rangle = \mathrm{Tr}_s \big\{\hat{\mathcal{P}}_{dark,N,s}\big\}\\
&=&
N_{dark,N,s}.
\label{eq.p2}\end{eqnarray}
We have used the fact that $\vert \psi_k\rangle$'s are basis states that span the $s$-sector.
In the last step, we have replaced the sum by a trace of $\mathcal{P}_{dark,N,s}$ over the $s$-sector. As $\mathcal{P}_{dark,N,s}$ is a projection operator, this operation directly yields the number of independent dark states.

We have demonstrated that the number of dark states (the dimensionality of the dark subspace) can be directly measured using a suitably designed experiment. It is noteworthy that this quantity is robust to arbitrary levels of disorder in the qubit-photon coupling constants, as shown in Sec.~\ref{sec.robust} above. 

\section{Dark-Bright Phase transition}
\label{sec.transition}
A phase transition requires three notions: a tuning parameter, an order parameter and a definition of the thermodynamic limit.  
As the tuning parameter, we propose
\begin{eqnarray}
 \alpha_{s,N}=s/N,
 \end{eqnarray}
which represents the fraction of excited qubits in the initial state. It  can also be viewed as the polarization of the initial state. Note that $\alpha_{s,N}$ can take values from zero to unity.

As the order parameter, we propose
\begin{eqnarray}
 o_{s,N}&=&
 N_{dark,N,s}/\left(
\begin{array}{c}
N\\
s
\end{array}
\right)
%}
.
\end{eqnarray}
which represents the fraction of dark states in the $s$-sector. The numerator, $N_{dark,N,s}$, represents the number of dark states -- a measurable quantity as shown in Sec.~\ref{sec.nullity} above. The denominator represents the total number of states within the $s$-sector. 
Using the expression for $N_{dark,N,s}$ in Eq.~\ref{eq.Ndark_clean}, we may write
\begin{eqnarray}
    o_{s,N} = \left\{ \begin{array}{cc}
0, & \mathrm{if}~\alpha_{s,N}>1/2,\\
\frac{N-2s+1}{N-s+1}, & \mathrm{if}~\alpha_{s,N}\leq 1/2.\\
\end{array}
\right.
\end{eqnarray}
As the thermodynamic limit, we propose taking the limit $N\rightarrow\infty$, while preserving the ratio $\alpha=s/N$. We obtain
\begin{eqnarray}
    o_{N\rightarrow\infty} = \left\{ \begin{array}{cc}
0, & \mathrm{if}~\alpha>1/2,\\
\frac{1-2\alpha}{1-\alpha}, & \mathrm{if}~\alpha\leq 1/2.\\
\end{array}
\right.
\end{eqnarray}
where $\alpha$ represents the tuning parameter in the thermodynamic limit (with subscripts $s,N$ suppressed).

This expression matches the profile of an order parameter at a continuous phase transition. An `ordered' phase appears for $\alpha<1/2$, where $o_{N\rightarrow\infty}$ takes a finite value. This indicates a `dark' phase where there is a non-zero likelihood that no photons will be emitted. This probability may be substantial, approaching unity for $\alpha\rightarrow 0$. 
A `disordered' phase appears for $\alpha>1/2$ where $o_{N\rightarrow\infty}$ vanishes. This indicates a `bright' phase where photon emission is inevitable as there are no dark states. The critical point corresponds to $\alpha=1/2$, where the order parameter goes to zero. Note that the order parameter itself varies continuously, but its derivative is singular at the transition. This is a characteristic feature of second-order phase transitions. 

\begin{figure}
\includegraphics[width=3.4in]{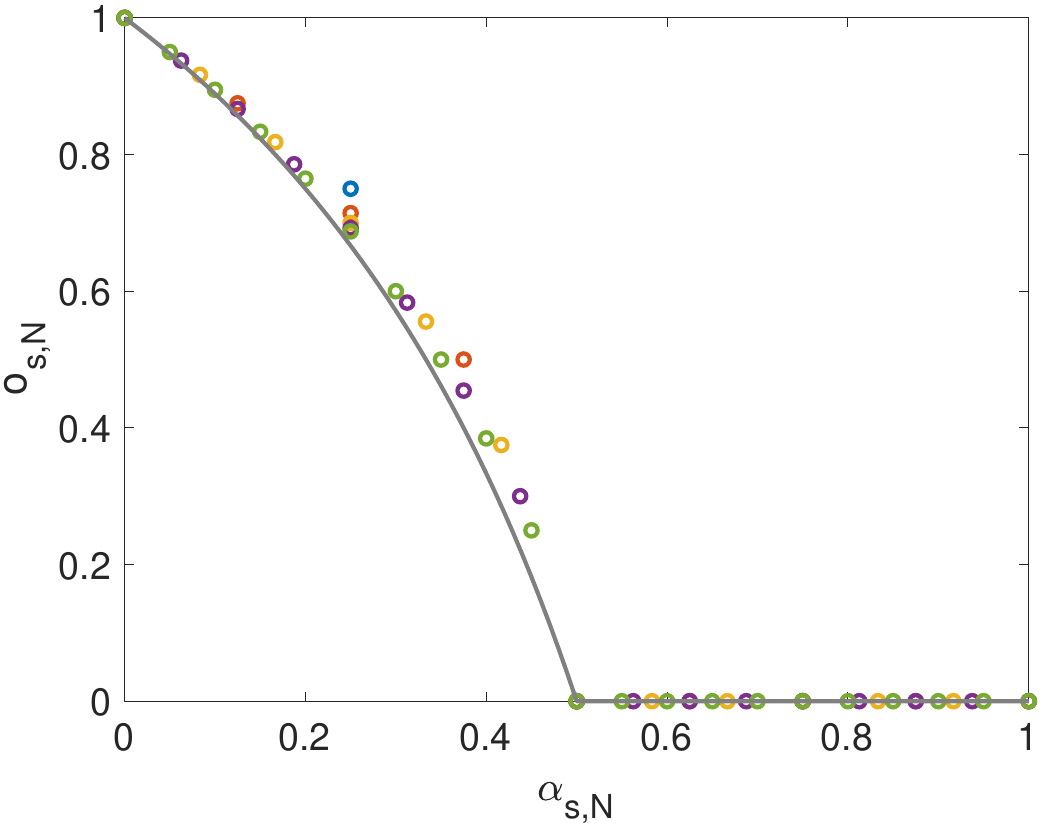}
\caption{The dark-to-bright phase transition. On the X axis, we have the tuning parameter $\alpha_{s,N}$, which corresponds to varying the polarization of initial states. The Y axis plots $o_{s,N}$, the fraction of dark states. The markers show values obtained for $N=4,8,12,16,20$. The solid line is the result expected in the $N\rightarrow \infty$ limit. }
\label{fig.phasetransition}
\end{figure}

In Fig.~\ref{fig.phasetransition}, we plot the order parameter as a function of the tuning parameter. The plot shows data for a few values of $N$, ranging from 4 to 20. It also shows the result expected in the thermodynamic $N\rightarrow \infty$ limit. A clear second-order phase transition emerges, with small finite-size corrections. 
Crucially, this phase transition is completely insensitive to disorder in the qubit-photon coupling constants ($g_j$'s). The order parameter and its behaviour do  not vary as $g_j$'s are varied.

\section{Discussion}
Our central result is an experimental protocol to measure the number of dark states in a Tavis-Cummings setup. Implementing this protocol in experiments can pose several challenges. The first is the need for a perfect photon detector. While the required technology exists at visible frequencies\cite{Wang2025}, efforts are underway to achieve perfect detection in the microwave regime\cite{Inomata2016,Khan2020,Pankratov2022,Chatterjee2023}. The second is the requirement to create direct-product states with various arrangements of excited- and ground-state-qubits. This can be achieved in circuit-QED setups with a few qubits. Even if all rearrangements cannot be accessed, a limited number of trials could set a lower bound for the number of dark states. The third involves effects beyond the standard Tavis-Cummings model such as inter-qubit coupling\cite{Gonzalez2016}, dephasing\cite{Davidsson2023}, counter-rotating terms\cite{Agarwal2012}, etc.

We have demonstrated that our results are completely insensitive to disorder in the qubit-photon coupling constants. In realistic setups, disorder may also appear in the qubit energies. The effect of such disorder on photon-number-dynamics has been studied\cite{Sun2022}. 
The effect on dark states has also been studied, but only within the single-excitation sector\cite{Zeb2022}. Future studies could examine their role in all sectors.

The key motivation for studying dark states is that they may serve as quantum memories. At the same time, dark character makes it difficult to produce or to manipulate such states. 
Our result provides a way to experimentally determine the number of dark states. This can pave the way for utilizing a Tavis-Cummings setup as a register with $N_{dark}$ possible states. Future studies can explore ways to create each state dark individually, as well to transform a given dark state to another.

\acknowledgments
LT thanks Pachaiyappan N for helpful discussions. RG thanks IIT Madras for warm hospitality and acknowledges support from National Sciences and Engineering Research Council of Canada (NSERC) through Discovery Grant 2022-05240. RN and LT acknowledge funding from the Center for Quantum Information Theory in Matter and Space-time, IIT Madras, and from the Department of Science and Technology, Government of India, under Grant No. DST/ICPS/QuST/Theme-3/2019/Q69, as well as support from the Mphasis F1 Foundation via the Centre for Quantum Information, Communication, and Computing (CQuICC).

\appendix
\section{Number of dark states in the uniform limit}
\label{app.Ndark_clean}
With our system of $N$ qubits, consider the space of states where $s$ qubits are excited while $(N-s)$ are in the ground state. All states in this space have $m_{(tot.)} = s/2-(N-s)/2 = s- N/2$. For any of these states to be `dark', we must have $S_{(tot.)} = N/2-s$ so that $m_{(tot.)} = -S_{(tot.)}$. This is not possible for $s>N/2$, as $S_{(tot.)}$ cannot be negative. Below, we focus on $s\leq N/2$ and determine the number of dark states.
 
The space with $s$ qubits excited is spanned by $\left( \begin{array}{c} N \\ s \end{array}\right)$ elements. The quantum number $S_{(tot.)}$ can take any of the following values: $N/2-s, N/2-s+1,\ldots,N/2$. We seek to exclude those with $(S_{(tot.)}>N/2-s)$ so that we are left with the dark states in this sector. We may do this as follows. Consider the sector with $s-1$ qubits excited while  $(N-s+1)$ are in the ground state. These states have $m_{(tot.)}=
s-1-N/2$, a negative value. They have $S_{(tot.)}=
N/2-s+1, N/2-s+2,\ldots,N/2$. Upon acting the $S_{(tot.)}^+$ operator on any of these states, we obtain states with $S_{(tot.)}>N/2-s$ but with $m_{(tot.)} = s-N/2$. These are precisely the states that must be excluded from the sector with $s$ qubits excited. The number of states to be excluded is given by $\left( \begin{array}{c} N \\ s-1 \end{array}\right)$. We conclude that the number of dark states in the sector with $s$ qubits excited is given by Eq.~\ref{eq.Ndark_clean} in the main text.

\section{Rank of $\hat{O}_s$}
\label{app.rank}
As defined in the main text, $\hat{O}_s$ is a matrix of dimensions $\left(\begin{array}{c} N\\ s -1 \end{array}\right)\times \left(\begin{array}{c} N\\ s\end{array}\right)$. We first consider the case $s \leq N/2$, where we have fewer rows than columns. The rank of $\hat{O}_s$ has an upper bound of $\left(\begin{array}{c} N\\ s -1 \end{array}\right)$.
We label the basis vectors of the source space as $\vert s_j\rangle$, where $j=1,\ldots,\left(\begin{array}{c} N\\ s \end{array}\right)$. These represent direct-product-states with $s$ qubits excited and $(N-s)$ qubits in the ground state. We represent the target states as $\vert t_k\rangle$, where $k=1,\ldots,\left(\begin{array}{c} N\\ s-1 \end{array}\right)$. These states span the subspace with $(s-1)$ qubits excited. Denoting the entries in $\hat{O}_s$ as $a_{k,j}$, they are defined as 
\begin{eqnarray}
a_{k,j} = \langle t_k \vert \left(\sum_{i} g_i S_{i}^{-}\right) \vert s_j \rangle.
\end{eqnarray}
If the rows in $\hat{O}_s$ are linearly dependent, there must exist $\{b_k \}$ such that 
\begin{eqnarray}
    \sum_k b_k a_{k,j} = 0,
\end{eqnarray}
for every $j$. Using the definition of $a_{k,j}$, we have
\begin{eqnarray}
    \sum_k b_k \langle t_k \vert \left(\sum_{i} g_i S_{i}^{-}\right) \vert s_j \rangle = 0,
\end{eqnarray}
Taking the complex conjugate, we have
\begin{eqnarray}
    \langle s_j \vert
     \left(\sum_{i} g_i^* S_{i}^{+}\right)  
     \left\{\sum_k b_k^* \vert t_k \rangle \right\} 
     = 0,
\end{eqnarray}
This equation must hold for any $j$. As $\vert s_j\rangle$'s are independent basis vectors for the $s$-subspace, this is only possible if  
\begin{eqnarray}
     \left(\sum_{i} g_i^* S_{i}^{+}\right)  
     \left\{\sum_k b_k^* \vert t_k \rangle \right\} 
     = 0.
     \label{eq.zero}
\end{eqnarray}
We argue that this equation can only hold true if every $b_k$ vanishes. In other words, the rows of $\hat{O}_s$ must be linearly independent. To see this, we observe that the raising operators $S_{i}^+$ act on $\vert t_k \rangle$'s to produce states with $s$ qubits excited ($\vert s_{j}\rangle$'s). As every raising operator acts on the $\vert t_k \rangle$'s, their action will produce all $\vert s_j \rangle$'s, which number $\left(\begin{array}{c} N\\ s \end{array}\right)$. This is greater than the number of free parameters, $b_k$'s, which is given by
$\left(\begin{array}{c} N\\ s-1 \end{array}\right)$. As a result, Eq.~\ref{eq.zero} cannot have a non-trivial solution. All $b_k$'s must be zero, showing that the rows of $\hat{O}_s$ are linearly independent. It follows that $\hat{O}_s$ has maximal rank, given by $\left(\begin{array}{c} N\\ s-1 \end{array}\right)$ for $s \leq N/2$.

We next consider $s>N/2$, with fewer columns than rows. In this case, the rank of $\hat{O}_s$ has an upper bound given by $\left(\begin{array}{c} N\\ s \end{array}\right)$. Suppose the columns are linearly dependent, we must have $\{c_j\}$ such that
\begin{eqnarray}
    \sum_j a_{k,j} c_j= 0
\end{eqnarray}
must hold for all $k$. That is, 
\begin{eqnarray}
    \langle t_k \vert \left(\sum_{i} g_i S_{i}^{-}\right) \left\{\sum_j  c_ j\vert s_j \rangle \right\}= 0.
\end{eqnarray}
This can only be possible if 
\begin{eqnarray}
    \left(\sum_{i} g_i S_{i}^{-}\right) \left\{\sum_j  c_ j\vert s_j \rangle \right\}= 0.
    \label{eq.zero2}
\end{eqnarray}
Here, the action of ${S}_i^-$ operators on the $\vert s_j\rangle$'s will produce all states of the target space. This corresponds to $\left(\begin{array}{c} N\\ s-1 \end{array}\right)$ independent states. This exceeds the number of free parameters, $c_j$'s, in Eq.~\ref{eq.zero2}, which is $\left(\begin{array}{c} N\\ s \end{array}\right)$. It follows that all $c_j$'s must vanish, indicating that the columns are linearly independent. The matrix $\hat{O}_s$ has rank $\left(\begin{array}{c} N\\ s \end{array}\right)$ when $s>N/2$.

\nocite{*}

\bibliography{countdark}

\end{document}